# Thermodynamic interpretation of the scaling of the dynamics of supercooled liquids

R.Casalini[a,b], U.Mohanty[c] and C.M.Roland[a]

[a] Naval Research Laboratory, Chemistry Division, Washington DC  20375-5342
[b] George Mason University, Chemistry Department, Fairfax VA  22030
[c] Boston College, Department of Chemistry, Newton, MA 02467

(April 25, 2006)

**ABSTRACT**

The recently discovered scaling law for the relaxation times, $\tau(T,\upsilon) = \Im(T\upsilon^\gamma)$, where T is temperature and $\upsilon$ the specific volume, is derived by a revision of the entropy model of the glass transition dynamics originally proposed by Avramov [I. Avramov, J. Non-Cryst. Solids **262**, 258 (2000).]. In this modification the entropy is calculated by an alternative route, while retaining the approximation that the heat capacity is constant with $T$ and $P$. The resulting expression for the variation of the relaxation time with $T$ and $\upsilon$ is shown to accurately fit experimental data for several glass-forming liquids and polymers over an extended range encompassing the dynamic crossover. From this analysis, which is valid for any model in which the relaxation time is a function of the entropy. we find that the scaling exponent $\gamma$ can be identified with the Grüneisen constant.

**I. INTRODUCTION**

Although the differences between the macroscopic properties of a liquid and solid are manifest, microscopically the two states are not so easily distinguished. The formation of a glass by progressive cooling (or compression) of a liquid is associated with a characteristic timescale for the dynamics, with the microscopic structure of the liquid retained. Given the ubiquitous presence of glassy materials in nature and their central importance to technologies in diverse fields such as biology, engineering, and geophysics, it is unsurprising that much effort is devoted to studying the glass transition. What might be surprising, however, is that research into this complex phenomenon remains at the model-building stage, with even the correct approach for the latter a contentious issue. Very generally, there are two interpretations based either on "free volume" concepts, whereby molecular motions are jammed in accord with the available unoccupied space [1,2], or on activated dynamics, with molecules transiently trapped in potential wells on the energy landscape [3,4,5]. While these models have been tested using many experimental techniques [6,7], the measurements usually involve temperature variations at atmospheric pressure. Such experiments convolute changes in density with changes in thermal energy, making difficult the identification of the factors governing the supercooled dynamics. Less often, due to experimental complexities, measurements are carried out as a function of hydrostatic pressure. These allow decoupling of volume and temperature effects, providing more rigorous tests of models for the glass transition.

Specific techniques for measuring relaxation of glass-forming liquids under high pressure include neutron scattering[8,9,10,11], light scattering[12,13,14,15,16,17,18,19], viscosity[20,21,22,23,24,25,26,27] and dielectric relaxation[28,29]. The latter has the advantage of providing a broad frequency range (routinely ten decades and even more at ambient pressure), which is essential since relaxation times vary by many orders of magnitude in the supercooled regime. Although dielectric spectroscopy measurements at elevated pressure were carried out forty years ago[30,31,32,33,34,35,36,37,38,39], there has been a bit of a lull until very recently. For a comprehensive review of high pressure measurements see ref. 28.

An important recent finding from high relaxation measurements of the dielectric relaxation time, $\tau$, is the existence of a scaling relation [40,41,42,43]

$$\tau(T,\upsilon) = \Im(T\upsilon^\gamma) \tag{1}$$

where $\upsilon(T,P)$ is the specific volume and $\gamma$ a material constant. This exponent is found to have values between 0.16 and 8.5 [40,41,42,43,44]. The power-law form enables accurate superpositioning over a broad range of $T$ and $\upsilon$. Alternate forms, such as a linear scaling of Tarjus and coworkers [45], have been proposed, but they fail for data encompassing an extended range [28,43]. Among possible justifications for the scaling, one hypothesis is that the repulsive part of the potential dominates the local liquid structure [46,47], so that for local properties the potential energy can be approximated with the spherically symmetric form [48,49]

$$U(r) = \varepsilon\left(\frac{\sigma}{r}\right)^n - \frac{a}{r^3} \tag{2}$$

where $\varepsilon$ and $\sigma$ are the characteristic energy and length scale of the system, $r$ is the intermolecular distance, and $n$ is a constant. The mean-field parameter $a$ describes the long-range attractive potential, which can be taken as a constant. Recent simulations of the glass transition have employed this inverse power repulsive potential[50,51]. A potential of this form suggests that the local properties should scale as a power law in $n$, or in terms of the volume according to eq.(1) for the dynamics (note $\gamma = n/3$). The implied scaling was verified for the Lennard-Jones 6-12 ($\gamma = 4$) liquid $o$-terphenyl (OTP) [11] by neutron scattering measurement, and subsequently over a broader range of frequencies by light scattering [17] and viscosity measurements [45].

For other materials the potential deviates from $\gamma = 4$ but by taking the exponent to be an adjustable parameter (but independent of T, $\upsilon$, and P), the scaling can be extended to a large number of supercooled liquids and polymers [40,41,42,43,44]. The wide range of values determined empirically for $\gamma$ makes problematic a direct connection between $\gamma$ and the repulsive potential exponent $n$ in eq.(2). Certainly the assumption of spherical symmetry cannot be strictly valid for interactions such as hydrogen bonds or the intramolecular bonds of a polymer backbone. Nevertheless, even in these cases the power law scaling of eq.(1) yields accurate superpositioning of relaxation times measured over a wide range of T and P.

The exponent $\gamma$ is a material constant determined by superpositioning of experimental data. The function $\Im(T\upsilon^{\gamma})$ is unspecified but for a given class of materials, e.g., organic liquids and polymers, is expected to have the same form. In this paper we eschew the intermolecular potential approach to $T\upsilon^{\gamma}$ scaling, adopting an alternative interpretation, based on an entropy model originally proposed by Avramov[52,53,54]. From an equation for the structural relaxation time (or viscosity) in terms of the configurational entropy, we derive a new expression for $\tau(T,\upsilon)$. This equation is found to describe the relaxation times over the entire frequency range, while satisfying the scaling (eq.(1)). Furthermore this equation for $\tau(T,\upsilon)$ is valid for a range of glass-forming materials. Of greater significance, we derive the parameter $\gamma$ in terms of thermodynamical quantities, thus conferring a more general identification of the $T\upsilon^{\gamma}$-scaling behavior. Finally we show that the scaling (eq.(1)) is not limited only to the Avramov model, but is valid for any model in which the relaxation time is governed by the entropy (such as the fluctuation model described herein in the appendix).

## II. THE MODEL

The Avramov model [52,53] is based on the notion that molecular motions are thermally activated, with a jump frequency given by

$$v_i(E_i) = \exp\left(-\frac{E_i}{RT}\right) \qquad (3)$$

Structural disorder gives rise to cooperative motion surmounting a broad distribution of barrier heights, whose mean frequency is

$$\langle v \rangle = \int_0^{E_{max}} v(E)\varphi(E,\sigma)dE \qquad (4)$$

where $\varphi(E,\sigma)$ is the probability of barrier energy $E$ and $\sigma$ is the variance of the Poisson distribution

$$\varphi(E,\sigma) = \frac{\exp\left[(E-E_{max})/\sigma\right]}{\sigma(1-\exp(-E_{max}/\sigma))} \qquad (5)$$

The details of $\varphi(E,\sigma)$ are unimportant, since the maximum of eq.(4) corresponds to values of $E_i$ far from its maximum [53]. From eqs.(3)-(5)

$$\langle v \rangle \cong v_0 \exp\left(-\frac{E_{max}}{\sigma}\right) \qquad (6)$$

where $v_0$ is a constant. This indicates that the dynamics is governed primarily by changes of the dispersion $\sigma$ rather than of the barrier height.

The entropy ($S$) and $\sigma$ are related according to [55]

$$\sigma = \sigma_r \exp\left[\frac{2(S-S_r)}{ZR}\right] \qquad (7)$$

where $\sigma_r$ is the dispersion at a reference state with entropy $S_r$, and $Z$ is the degeneracy of the system; i.e., the number of available pathways for local motion of a molecule or polymer segment (roughly proportional to the coordination number of the liquid lattice). From eqs.(6) and (7) it follows that

$$\tau = \tau_0 \exp\left\{\varepsilon \exp\left[-\frac{2(S-S_r)}{ZR}\right]\right\} \qquad (8)$$

where $\varepsilon = E_{max}/\sigma_r$ and $\tau_0$ is the limiting value at high temperatures.

It can be seen that according to the Avramov model, the behavior of $\tau$ (or $\eta$) is mainly a function of the entropy, while $\varepsilon$ and $\tau_0$ are considered constants. The temperature dependence of the relaxation time at atmospheric pressure is then obtained by calculating the entropy $S(T)$ using the approximation that $C_P$ (of the equilibrium liquid) is temperature independent

$$S(T) = S_r + \int_{T_r}^{T} C_P d\ln T' = S_r + C_P \ln\left(\frac{T}{T_r}\right) \quad (9)$$

Substituting into eq.(8) gives

$$\tau(T) = \tau_0 \exp\left\{\varepsilon\left(\frac{T_r}{T}\right)^{\frac{2C_P}{ZR}}\right\} \quad (10)$$

This equation has been found to accurately describe experimental data over a wide dynamic range[53].

Extending the model to high pressure, the entropy is calculated as a function of $T$ and $P$ [53,56]

$$\tau(T,P) = \tau_0 \exp\left\{\varepsilon\left(\frac{T_r}{T}\right)^{\frac{2C_P}{ZR}}\left(1+\frac{P}{\Pi}\right)^{\frac{2\alpha_P V_m \Pi}{ZR}}\right\} \quad (11)$$

where $\alpha_P$ $(=V^{-1}(\partial V/\partial T)_P)$ is the isobaric thermal expansion coefficient at atmospheric pressure, $V_m$ is the molar volume and $\Pi$ is a constant. To calculate eq.(11) from eq.(8), it is assumed that $\alpha_p$ is inversely proportional to $P$. This expression gives a satisfactory description of experimental $\tau(T,P)$ data and also yields an expression for the pressure-dependence of the glass transition temperature identical to the empirical Andersson equation[57,58], the latter widely used to fit $T_g(P)$ results. (The Andersson equation can also be derived from the Simon equation[59,76].) One shortcoming, however, is that the value of the Avramov parameters calculated from thermodynamic quantities can differ from the values obtained by fitting of experimental relaxation times.[58] Another problem is that according to eq.(11), the steepness index (or fragility) [4,60,61] defined as

$$m_P = \frac{1}{T_g}\left[\frac{\partial \log_{10}(\tau)}{\partial(1/T)}\right]_{T=T_g} \quad (12)$$

is a constant independent of $P$. In fact, experimental results show unambiguously that $m_P$ decreases with increasing $P$ [28,62,63,64].

Herein we use the Avramov eq.(8) as our starting point, but adopt a different approach for calculation of the entropy, using

$$dS = \left.\frac{\partial S}{\partial T}\right|_V dT + \left.\frac{\partial S}{\partial V}\right|_T dV = \frac{C_V}{T}dT + \left.\frac{dP}{dT}\right|_V dV = \frac{C_V}{T}dT + \frac{C_P - C_V}{V\alpha_P T}dV \qquad (13)$$

where the expression for $\partial P/\partial T|_V$ follows from the thermodynamic relationship

$$C_P = C_V + \frac{TV\alpha_P^2}{\kappa_T} = C_V + TV\alpha_P \left(\frac{\partial P}{\partial T}\right)_V \qquad (14)$$

with $\kappa_T (= -1/V\, \partial V/\partial P|_T)$ the isothermal compressibility. Using the fact that $S$ is a function of state and considering $C_V$ to be constant with respect to $T$ and the difference $C_P - C_V$ to be constant with respect to $V$ (as in the original model and approximately true over modest ranges of $T$ and $P$ – these approximations and their consequences are discussed at the end of the following section), it follows that

$$S(T,\upsilon) = S_r + C_V \left[ \ln\left(\frac{T}{T_r}\right) + \frac{C_P/C_V - 1}{\alpha_P T} \ln\left(\frac{\upsilon}{\upsilon_r}\right) \right] \qquad (15)$$

Defining

$$\gamma = \frac{C_P/C_V - 1}{\alpha_P T} \qquad (16)$$

we obtain

$$S(T,\upsilon) = S_r + C_V \ln\left(\frac{T\upsilon^\gamma}{T_r \upsilon_r^\gamma}\right) \qquad (17)$$

From the model (eq.(8)) the relaxation time is then given by

$$\tau(T,\upsilon) = \tau_0 \exp\left[\varepsilon\left(\frac{T_r \upsilon_r^\gamma}{T\upsilon^\gamma}\right)^{\frac{2C_V}{ZR}}\right] \qquad (18)$$

Using

$$D = \frac{2C_V}{ZR} \qquad (19)$$

eq.(18) can be rewritten as

$$\tau(T,\upsilon) = \tau_0 \exp\left[\varepsilon\left(\frac{T_r \upsilon_r^{\gamma_G}}{T \upsilon^{\gamma_G}}\right)^D\right] \qquad (20)$$

In eq.(20) the parameters $\tau_0$, $\varepsilon$, $\gamma_G$ and $D$, as well as the reference temperature $T_r$ and volume $\upsilon_r$ are constants; therefore, this function satisfies the scaling relation (eq.(1)). However, it remains to be demonstrated whether eq.(20) provides a satisfactory description of $\tau(T,\upsilon)$ data. In the following section we fit this equation to experimental relaxation times for various glass-formers.

Eq.(20) is more general than the Avramov model underlying the above derivation. As shown in the appendix, the same expression can be obtained within the framework of fluctuation theory, thus establishing a firm basis for the $T\upsilon^\gamma$-scaling in an entropy conception of the glass transition dynamics.

### III. TEST OF THE MODEL

To fit eq.(20) to experimental data, we rewrite it as

$$\log\left[\tau(T,\upsilon)\right] = \log(\tau_0) + \left(\frac{B}{T\upsilon^\gamma}\right)^D \qquad (21)$$

Therefore there are four parameters available to describe measurements for all $T$ and $\upsilon$, one more than required to fit only isobaric data (eq.(10)) and one less than eq.(11) of the original Avramov model for elevated pressure [52,56]. Employing the common definition that the relaxation time at the glass temperature $\tau(T_g,\upsilon_g)=100s$, and taking the ambient pressures values of $T_r=T_g$ and $\upsilon_r=\upsilon_g$, it follows that $\varepsilon = \ln(100/\tau_0)$ and therefore $B = \varepsilon^{\frac{1}{D}}\left(T_g V_g^{\gamma_G}\right)$. This reduces the number of adjustable parameters in eq.(21) to three ($\gamma_G$, $D$, $\tau_0$).

The best fits obtained for 1,1'-di(4-methoxy-5-methylphenyl)cyclohexane (BMMPC)[65,66], 1,2-polybutadiene (1,2-PB) [67], phenolphthalein-dimethyl-ether (PDE) [68,69], D-sorbitol [70], propylene carbonate (PC) [71], and polymethylphenylsiloxane (PMPS) [72] are displayed in Figs.1 through 6, respectively, which show isobars and isotherms as a function of specific volume. Also shown as an inset are Arrhenius plots of the ambient-pressure data. These particular glass-formers were chosen because they represent a range of dynamic behavior, as evidenced by the range of the scaling exponent, $0.16 \leq \gamma \leq 8.5$ [40,41,42,43]. For each material, the fitting was carried out simultaneously on all data sets; the obtained parameters and the statistical significance of the fit ($\chi^2$) are listed in table 1. The functional form $\tau(T,\upsilon)$ (eq.(21)) derived from the Avramov model describes the data very well over the entire range for the different thermodynamical conditions.

Calculating the isobaric fragility $m_P$ from eq.(20), with the reference temperature taken as the glass temperature $T_r=T_g$ and with $\gamma_G$ and $D$ constant, we obtain

$$m_P = \varepsilon D(1+\alpha_P T_g \gamma_G) \qquad (22)$$

This expression, unlike the equation derived originally in the extension of the Avramov model to high temperature [52,56], does not predict an invariance of $m_P$ to pressure, since both $\alpha_P$ and $T_g$ are pressure dependent and their product decreases with P [64]. The present prediction that $m_P$ decreases with pressure agrees with the general experimental result for non-associated glass-formers, $dm_P/dP < 0$. [64] (Associated liquids, e.g., water, for which $\alpha_P<0$ or $\alpha_P=0$ may exhibit other behavior. Since the equation of the original model is the same as eq.(8), the difference between the predicted $m_P(P)$ behavior is ascribed to the different approximations used to calculate $S$.

Similarly we can calculate the isochoric (constant volume) fragility, $m_V$ from

$$m_V = \varepsilon D \qquad (23)$$

This indicates $m_V$ is a constant, in agreement with experimental results [64]. The ratio of the two fragilities is given by

$$\frac{m_V}{m_P} = \frac{1}{1+\gamma_G \alpha_P T_g} \qquad (24)$$

a relation which has to be satisfied by any form of $\tau(T,\upsilon)$ satisfying the scaling relation $\tau(T,\upsilon) = \Im(T\upsilon^\gamma)$ [41, 64].

From the equations for $\gamma_G$ and $D$ it follows that

$$m_P = \frac{2\varepsilon C_P}{ZR} \qquad (25)$$

and

$$m_V = \frac{2\varepsilon C_V}{ZR} \qquad (26)$$

which together with eq.(14) give

$$m_P = m_V + \frac{2\varepsilon}{ZR}\frac{TV\alpha_P^2}{\kappa_T}\bigg|_{T=T_g} \qquad (27)$$

Taking $\tau(T_g,\upsilon_g)=100s$

$$m_P = m_V + \frac{2\ln(100/\tau_0)}{ZR}\frac{TV\alpha_P^2}{\kappa_T}\bigg|_{T=T_g} \tag{28}$$

This form resembles the recently reported linear correlation between the isobaric and isochoric fragilities[73]

$$m_P = (0.84 \pm 0.05)m_V + (37 \pm 3) \tag{29}$$

Comparing these two equations, we note that $m_P > m_V$, which means that $\tau$ cannot be a function only of $T$ (since that requires $m_P = m_V$).

In table 2 we list the value of Z calculated by equating the second term of eq.(28) to its empirically-determined value of 37 [73] (together with other known thermodynamic properties). We find that the parameter Z varies herein from 1.9 for 1,2-PB to 17.8 for BMMPC. This is in accord with the original work of Avramov[74], who suggested Z~2 for polymers and Z~10 for molecular liquids.

From the definition of $\gamma_G$ (eq. (16)), together with eq.(14), we can express $\gamma_G$ as

$$\gamma_G = \frac{V\alpha_P}{C_V \kappa_T} \tag{30}$$

This is the well-known thermodynamic definition of the Grüneisen constant[75]. However, the identification of the scaling exponent $\gamma$ with $\gamma_G$, is not trivial[76]. The Grüneisen constant is generally defined in terms of the change of the vibrational frequency with volume[75], which means that it includes thermodynamic contributions related to different degrees of freedom than those relevant to relaxation.

The result of eq.(30) is not limited to the entropy model considered herein. In fact, since eq.(30) can be rewritten as

$$\frac{V\alpha_P}{C_V \kappa_T} = (V/T)\frac{dS}{dV}\bigg|_T \bigg/ \frac{dS}{dT}\bigg|_V \tag{31}$$

the scaling exponent defined by eq.(1) is given by eq.(30) for any model for which $\tau$ is a function of $S(T,\upsilon)$, provided the entropy has the form

$$S(T,\upsilon) = \Im'(T\upsilon^{\gamma_G}) \tag{32}$$

This is valid for the expression for $S(T,\upsilon)$ obtained herein (eq.(17)). In fact, as recently shown[76], the result $\gamma \sim \gamma_G$ can also derived from the assumption of a power-law intermolecular potential in combination with eqs. (31) and (32).

In table 3 the parameter $\gamma_G$ is calculated using eq.(30) from literature data, with the isochoric heat capacity taken from its isobaric value using eq.(14). We find that the $\gamma_G$ obtained in this fashion are all smaller (by roughly one-third) than the values deduced from superpositioning of experimental relaxation data. This discrepancy between the predictions from the thermodynamic values and the direct experimental γ may be explained (i) in terms of the different contributions to the heat capacity, some of which may not affect τ, and (ii) the non ideal behavior of the thermal pressure coefficient, $\partial P/\partial T|_V$. Both these items are discussed in the following.

The heat capacity $C_P$ of the liquid involves contributions from other motions not involved in structural relaxation but their removal is not trivial. One approach (used in evaluating other entropy models[77]) is to use the difference, $\Delta C_P$, of the heat capacities of the liquid and crystal rather than $C_P$. We make the assumption that the ratio between $C_V$ and $C_P$ is equal to the ratio $\Delta C_V / \Delta C_P$. The exponent, $\gamma'_G$, calculated using this value for the heat capacity is in all cases (excepting PMMA) quite close to the value of γ obtained from the scaling of relaxation data (table 3).

Regarding the approximations used to calculate the entropy, in eq.(13) we considered $(C_P - C_V)/\alpha_P T$ to be independent of $\upsilon$ (which is equivalent to taking $\left.\frac{\partial P}{\partial T}\right|_\upsilon \propto \frac{1}{\upsilon}$). This approximation is not necessarily accurate even if $C_V$ and $C_P$ are approximately independent of $\upsilon$. To assess this approximation, we calculated $\left.\frac{\partial P}{\partial T}\right|_\upsilon$ from the parameters of the equation of state for different values of the volume. The values of the derivative for different materials are plotted versus the reciprocal of $\upsilon$ in Fig.7. The behavior can be well described by a linear equation $\left.\frac{\partial P}{\partial T}\right|_\upsilon = a + \frac{b}{\upsilon}$, for which the best-fit parameters (solid lines in Fig.7) are listed in Table 4. Since $a \sim b$, the $a$ cannot necessarily be neglected. If we include the linear expression for $\left.\frac{\partial P}{\partial T}\right|_\upsilon$ in eq.(13), the entropy relation becomes

$$S(T,\upsilon) = S_r + C_V \ln\left(\frac{T}{T_r}\right) + \bar{b} \ln\left(\frac{\upsilon}{\upsilon_r}\right) + \bar{a}(\upsilon - \upsilon_r) \quad (33)$$

where $\bar{a}$ and $\bar{b}$ are the parameters $a$ and $b$ multiplied by the molecular weight.

Since for the change of volume considered $\upsilon/\upsilon_r \sim 1$, this equation can be well-approximated as

$$S(T,\upsilon) = S_r + C_V \ln\left(\frac{T}{T_r}\right) + \left(\bar{b} + \bar{a}\upsilon_r\right)\ln\left(\frac{\upsilon}{\upsilon_r}\right) \qquad (34)$$

and therefore the expression for $\tau(T,\upsilon)$ becomes

$$\tau(T,\upsilon) = \tau_0 \exp\left[\varepsilon\left(\frac{T_r \upsilon_r^{\gamma_G}}{T \upsilon^{\gamma_G}}\right)^{\frac{2C_V}{ZR}}\right] \qquad (35)$$

where

$$\gamma_G = (\bar{b} + \bar{a}\upsilon_r)/C_V = (\bar{b} + \bar{a}\upsilon_r)/[C_P - T\upsilon\alpha_P(\bar{a} + \bar{b}/\upsilon)] \qquad (36)$$

From eq.(36) we see that $\gamma_G$ is dependent on $\upsilon$. However, since the term in the denominator depending on $\upsilon$ (i.e, the difference between $C_P$ and $C_V$) is much smaller than the first term, this dependence of $\gamma_G$ on $\upsilon$ is negligible: We estimate that the change of $\gamma_G$ is less than 5% over the entire range of the data considered herein. Moreover, eq.(36) can be rewritten as

$$\gamma_G = \frac{V_r}{C_V}\left.\frac{\partial P}{\partial T}\right|_{V=V_r} \qquad (37)$$

which is essentially equivalent to eq.(30) for small relative changes of volume. So the values of the parameter $\gamma_G$ are those reported in table 3, and in practice even if the volume dependence of the thermal pressure coefficient were not strictly proportional to the volume, in the range over which we have tested our model, the error is negligible.

**IV. Concluding Remarks**

The idea for the $T\upsilon^\gamma$ scaling[40,41] arises from consideration of a generalized repulsive potential[48,49], which is drawn from a Lennard-Jones type intermolecular potential[11,17,45]. Although the material-constant $\gamma$ is determined empirically, the function $\Im(T\upsilon^\gamma)$ itself is unknown *a priori*. Starting from an equation originally proposed by Avramov[74,83,84,85], which related the relaxation times of glass-formers to the entropy change accompanying vitrification (eq.(8)), the function $\tau = \Im(T\upsilon^\gamma)$ is derived with $\gamma \sim \gamma_G$. The difference in our approach from that of Avramov is the thermodynamic paths used to calculate the total entropy, whereby some of the approximations used in the original derivation are avoided. Moreover, obtaining an expression for $\tau(T,\upsilon)$ having the form of eq.(21) does not rely on the Avramov approach; it can be derived from any model in which the relaxation time is governed by the entropy, since the change of entropy is a function of $T\upsilon^\gamma$ (eq.(17))). This is demonstrated in the appendix, wherein the same expression for $\tau(T,\upsilon)$ is obtained from fluctuation theory.

The modified Avramov equation (eq.(21)) accounts well for the variation of relaxation times with $T$ and $\upsilon$ for a variety of organic liquids and polymers. Beyond the success of $\Im(T\upsilon^\gamma)$ as a fitting function for a broad range of experimental variables, the exponent $\gamma$ is now related to thermodynamic quantities, providing a new and more rigorous basis for the $T\upsilon^\gamma$-scaling. Specifically, we find that $\gamma$ can be identified with the Grüneisen parameter. The connection of the scaling exponent to other molecular properties is of particular interest because it suggests the possibility of extracting the $P$- and $\upsilon$-dependences of $\tau$ from measurements merely at ambient pressure. The main limitation appears to be obtaining reliable values for $\gamma_G$. This is related to the need to use only that part of the entropy related to the structural relaxation, as shown from the better agreement between $\gamma$ and $\gamma'_G$ where for the latter the difference of the heat capacity between the liquid and the crystal was used, rather than the heat capacity of the liquid.

The revised model gives the correct dependence of the fragility on pressure, in contrast to the original erroneous prediction[63] from the Avramov model that the fragility was independent of pressure. We also obtain a relationship between the isochoric and isobaric fragilities in quantitative accord with a recently reported empirical correlation (eq.(29))[73]. From this we estimate the number of available pathways for local motion; the results (Z~2 for polymers and Z~10 for small molecules) are consistent with expectations from the original Avramov model.

Commonly, the T-dependence of the dynamics of supercooled liquids at constant pressure is described using the Vogel-Fulcher-Tammann-Hesse (VFTH) function[86,87,88]. However, the VFTH function is limited to data above a characteristic ("dynamic crossover") relaxation time $\tau_B$ [69,89,90,91,92], whereas eq.(21) describes $\tau(T,\upsilon)$ over the entire dynamic range, including variations in P as well as T. (Note, however, all data herein are below the temperature at which the relaxation time assumes Arrhenius behavior. This temperature is significantly greater than the temperature of the dynamic crossover.[89]) Other models, such as that due to Adam and Gibbs (AG)[77] and its extension to high pressure[93,94,95,96,97,98,99], also fail to describe $\tau(T,P) < \tau_B$[95,100]. The only model found[101,102,103] to fit data over a range encompassing the dynamic crossover is the Cohen-Grest (CG) free volume model[104], in which the dynamic crossover is identified with the percolation of free volume [103]. The CG model employs five adjustable parameters (one more than eq.(21)) for $\tau(T,P)$; moreover, the physical plausibility of the obtained parameters has been questioned[105], in addition to any inherent difficulties with a purely free volume approach[106].

An important feature of the Avramov model, distinguishing it from the AG and CG models and from functional forms such as the VFTH, is that eq.(21) does not predict any divergence of $\tau$ with decreasing $T$ and/or $\upsilon$. There is only a monotonic, progressive slowing down of the dynamics. As described herein, this slowing down is driven by the increasing heterogeneity of the dynamics related to the increased dispersion of the energy barrier distribution. Of course, an absence of any divergence is implicit in the scaling relation eq.(1).

**ACKNOWLEDGEMENTS**

This work was supported by the Office of Naval Research. Inspiring conversations with S. Capaccioli are gratefully acknowledged.

**APPENDIX: FLUCTUATION THEORY**

The basic idea is that Landau-Lifshitz thermodynamic fluctuation theory[107] in conjunction with the notion of cooperatively rearranging region [77] in a supercooled liquid lead to the key result of the Avramov model, namely eq.(21). A cooperatively rearranging region is a subsystem that upon sufficient fluctuations can rearrange itself, independent of its environment, leading to viscous flow [77]. The subsystem together with the remaining part of the system constitutes a closed system.

The probability of a fluctuation in the closed system within the framework of Landau-Lifshitz fluctuation theory is $P \propto \exp(\Delta S_t / R)$, where $\Delta S_t$ is the entropy change on fluctuation of the entire system [107]. Denoting by $\Delta S$, $\Delta V$, and $\Delta E$ as the respective changes in entropy, volume and energy upon fluctuations in the subsystem, then the minimum work due to reversible changes in thermodynamic quantities of the subsystem is

$$w_{\min} = T\Delta S_t = -(\Delta E - T\Delta S + P\Delta V) \tag{A1}$$

The probability of a fluctuation in the subsystem is then [107]

$$P = \exp\left[-(\Delta E - T\Delta S + P\Delta V)/RT\right] \tag{A2}$$

Note that eq.(A2) is valid for large as well as for small fluctuations [107].

If fluctuations are small, and since the (internal) energy $E$ is a function of $S$ and $V$, $\Delta E$ can be expanded in a Taylor series to quadratic order. Substituting this expansion in eq.(A2), we obtain the probability of a fluctuation in the subsystem or a cooperative rearranging region [107]

$$P \propto \exp\left[(2T)^{-1}\left.\frac{\partial V}{\partial P}\right|_S - (\Delta S)^2/2RC_P\right] \tag{A3}$$

Assuming that the temperature dependence of first term is weak relative to the $\Delta S$ term, we absorb the former into a factor, $A$, which is weakly temperature-dependent relative to the exponential term

$$P = A\exp\left[-(\Delta S)^2/2RC_P\right] \tag{A4}$$

The relaxation time for cooperative rearrangement is inversely proportional to the transition probability for a cooperative rearrangement

$$\tau = \tau_O \exp((\Delta S)^2/2RC_P) \tag{A5}$$

where $\tau_O$ is approximately $A^{-1}$.

From eq. (17), the change in entropy is $\Delta S = -C_V \ln(B/Tv^{\gamma_G})$. Substituting this expression in eq.(A5), we obtained the desired result

$$\ln\tau = \ln\tau_O + (B/Tv^{\gamma_G})^{C_V/gR} \tag{A6}$$

where $g = C_P/C_V$. Eq.(A6) is identical in form to eq.(21) drawn from the Avramov model.

# Figure Captions

**Figure 1.** Relaxation times of 1,1'-di(4-methoxy-5-methylphenyl)cyclohexane [65,66] at constant P=0.1 MPa (open symbols) and isothermal conditions (solid symbols) at the indicated temperatures. The solid line is eq.(21) with the fit parameters listed in table 1. The inset shows the relaxation times at atmospheric pressure as a function of inverse temperature (open symbols) together with the best fit (solid line).

**Figure 2.** Relaxation times of 1,2-polybutadiene [67] measured at constant pressure (open symbols) and at constant temperature (solid symbols). The solid line is eq.(21) with the fit parameters listed in table 1. The inset shows the relaxation times at atmospheric pressure as a function of inverse temperature (open symbols) together with the best fit (solid line).

**Figure 3.** Relaxation times for phenylphthalein-dimethylether [68,69] versus the specific volume. The data were measured at constant pressure (open symbols) and at constant temperature (solid symbols). The solid line is the fit to all data using eq.(21), with the fit parameters listed in table 1. The inset shows the relaxation times at atmospheric pressure as a function of inverse temperature (open symbols) together with the best fit (solid line).

**Figure 4.** Relaxation times of D-sorbitol [70] versus specific volume for isobaric (open symbols) and isothermal (solid symbols) measurements. The solid line is eq.(21) with the fit parameters given in table 1. The inset shows the relaxation times at atmospheric pressure as a function of inverse temperature (open symbols) together with the best fit (solid line).

**Figure 5.** Relaxation times of propylene carbonate [71] versus specific volume for isobaric (open symbols) and isothermal (solid symbols) measurements. The solid line is eq.(21) with the fit parameters given in table 1. The inset shows the relaxation times at atmospheric pressure as a function of inverse temperature (open symbols) together with the best fit (solid line).

**Figure 6.** Relaxation times of polymethylphenylsiloxane [72] versus specific volume for isobaric (open symbols) and isothermal (solid symbols) measurements. The solid line is eq.(21) with the fit parameters given in table 1. The inset shows the relaxation times at atmospheric pressure as a function of inverse temperature (open symbols) together with the best fit (solid line).

**Figure 7.** Temperature derivative of the pressure at fixed volume as a function of the inverse volume (calculated from the equation of state). The solid lines are the linear fits (parameters in table 4).

**Table 1.** Avramov fit parameters.

| Material | $\log(\tau_0)$ | B | $\gamma_G$ | D | $\chi^2$ | $\gamma^*$ |
|---|---|---|---|---|---|---|
| BMMPC | -11.37±0.12 | 411±9 | 8.2±0.1 | 2.03±0.04 | 0.83 | 8.5 [40] |
| 1,2-PB | -7.71±0.06 | 353±2 | 1.89±0.01 | 7.76±0.13 | 0.4 | 1.9 [64] |
| PDE | -9.37±0.04 | 129.2±0.9 | 4.36±0.02 | 4.33±0.04 | 0.6 | 4.5 [64] |
| sorbitol | -9.40±0.24 | 326±4 | 0.13±.002 | 9.2±0.4 | 0.79 | 0.16 [40] |
| PC | -10.30±0.02 | 91.3±0.4 | 3.82±0.01 | 4.62±0.03 | 0.6 | 3.7 [71] |
| PMPS | -10.3±0.2 | 185±4 | 5.63±0.02 | 4.7±0.1 | 0.83 | 5.6 [28] |

*Literature values determined from superpositioning of experimental $\tau(T,\upsilon)$.

**Table 2.** Properties of the materials at $T_g$ and atmospheric pressure. For polymers the molar volume refers to the repeat unit. The parameter Z was calculated from eq.(28) and (29) as $Z = \dfrac{1}{37}\dfrac{2\ln(100/\tau_0)}{R}\dfrac{TV\alpha_P^2}{\kappa_T}\bigg|_{T=Tg}$.

| Material | $T_g$ [K] | $V_g$ [cm$^3$mol$^{-1}$] | $\alpha_P \times 10^4$ [K$^{-1}$] | $\kappa_T \times 10^4$ [MPa$^{-1}$] | Z |
|---|---|---|---|---|---|
| BMMPC[a] | 263 | 196.4 | 7.90 | 3.62 | 17.8 |
| 1,2 PB[b] | 253.5 | 56.44 | 7.10 | 5.50 | 1.9 |
| PDE[c] | 298 | 255.07 | 6.08 | 3.64 | 13.1 |
| sorbitol[d] | 267 | 111.58 | 4.45 | 1.14 | 8.8 |
| PC[e] | 158.3 | 77.16 | 6.72 | 2.14 | 4.7 |
| PMPS[f] | 245 | 118.2 | 5.80 | 3.6 | 4.9 |

[a] Ref. 66, [b] Ref. 67, [c] Ref. 18, [d] Ref. 70, [e.] Ref. 71, [f.] Ref. 72

**Table 3.** Comparison of the Grüneisen constant at $T \sim T_g$ ($T$ was chosen as close as possible to $T_g$ but avoiding interpolation of $C_P$ data) with the scaling exponent $\gamma_G$ calculated according to eq.(30) and $C_V$ obtained from eq.(14). $\gamma'_G$ was calculated using the difference $C_P^{liq}-C_P^{cryst}$ rather than $C_P^{liq}$ and assuming $C_P^{liq}/C_V^{liq}=(C_P^{liq}-C_P^{cryst})/(C_P^{liq}-C_P^{cryst})$.

| Material | $T$ [K] | $V_g$ [cm$^3$mol$^{-1}$] | $\alpha_P \times 10^4$ [K$^{-1}$] | $\kappa_T \times 10^4$ [MPa$^{-1}$] | $C_P^{liq}$ [Jmol$^{-1}$K$^{-1}$] | $C_P^{cryst}$ [Jmol$^{-1}$K$^{-1}$] | $\gamma_G$ | $\gamma'_G$ | $\gamma$ |
|---|---|---|---|---|---|---|---|---|---|
| OTP | 247 | 206.1[a] | 7.08[a] | 4.2[a] | 338.3[b] | 225.16[b] | 1.2 | 3.6 | 4 |
| PVAc | 304 | 72.50[c] | 7.15[c] | 5.0[c] | 156.91[d] | 116.2[d] | 0.7 | 2.7 | 2.5 |
| PMMA | 380 | 86.96[e] | 5.8[e] | 3.9[e] | 203.16[d] | 166[d] | 0.7 | 3.8 | 1.25 |
| salol | 220 | 169.25[f] | 7.85[f] | 3.09[f] | 298.41[g] | 186.7[g] | 1.9 | 5.1 | 5.2 |
| PC | 164 | 69.43[h] | 6.26[h] | 2.21[h] | 158.56[i] | 87.22[i] | 1.4 | 3.1 | 3.7 |

[a.] Ref. 44, [b.] Ref. 78, [c.] Ref. 58, [d.] Ref. 79, [e.] Ref. 80, [f.] Ref. 19, [g.] Ref. 81, [h.] Ref. 71, [i.] Ref. 82.

**Table 4.** Linear fit parameters of the dependence of $\partial P/\partial T|_v$ on inverse volume (solid lines in Figure 7).

| Material | $a$ [MPaK$^{-1}$] | $b$ [MPaK$^{-1}$mlg$^{-1}$] |
|---|---|---|
| OTP | -5.8±0.2 | 6.6±0.2 |
| PVAc | -6.1±0.2 | 6.3±0.2 |
| PMMA | -2.1±0.2 | 2.7±0.2 |
| salol | -10.3±0.2 | 10.0±0.2 |
| PC | -5.6±0.2 | 6.5±0.2 |

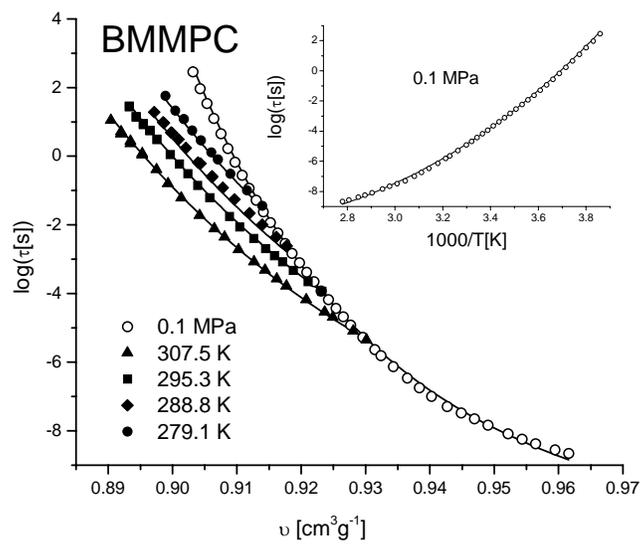

Figure 1

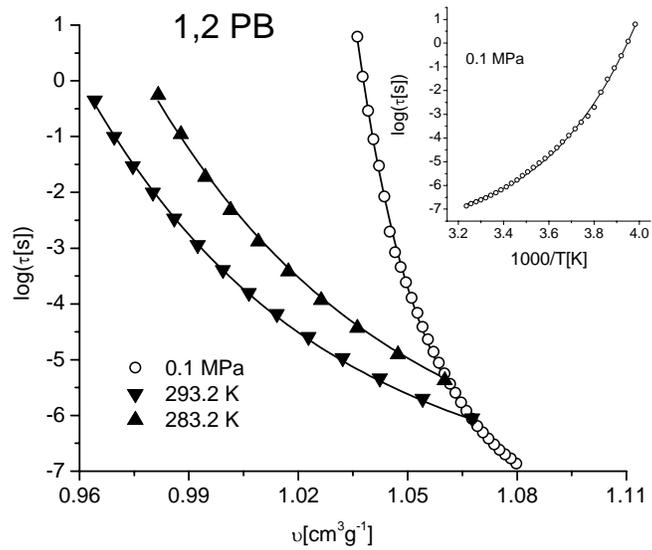

Figure 2

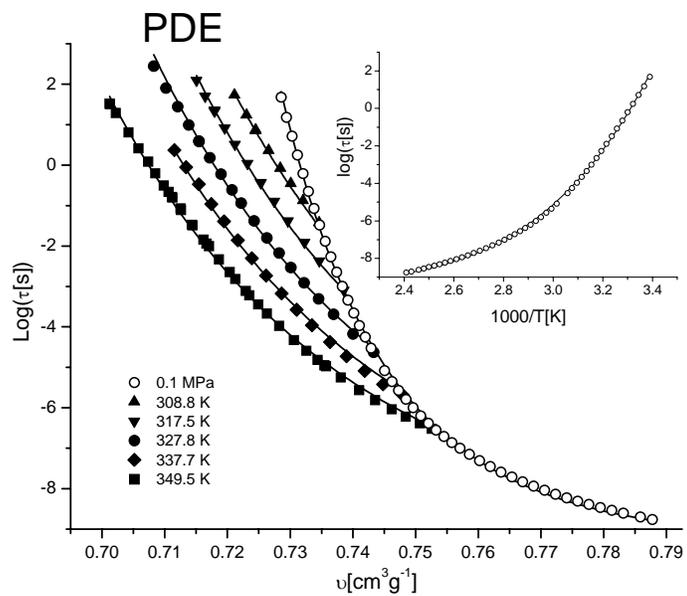

Figure 3

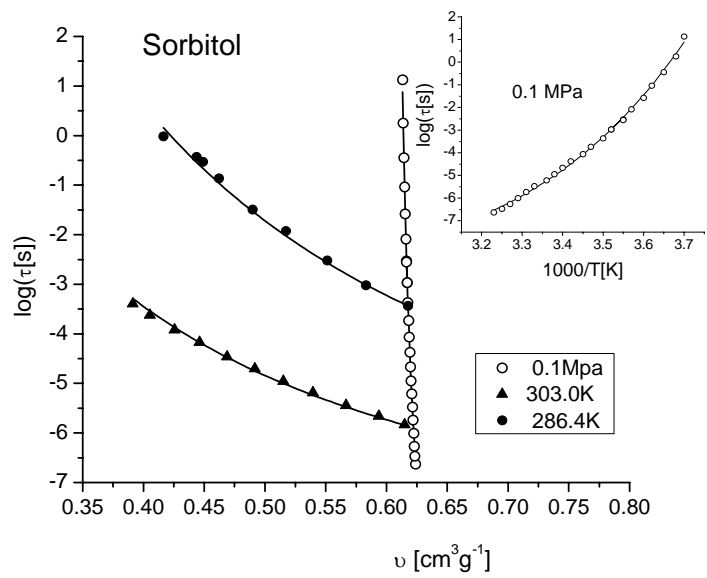

Figure 4

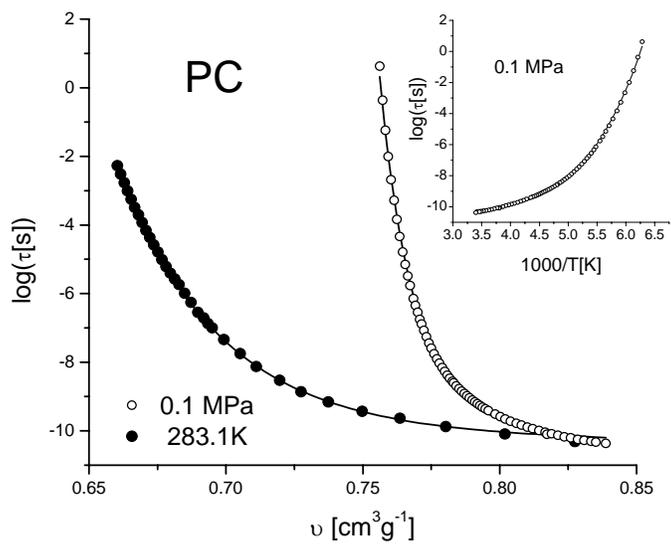

Figure 5

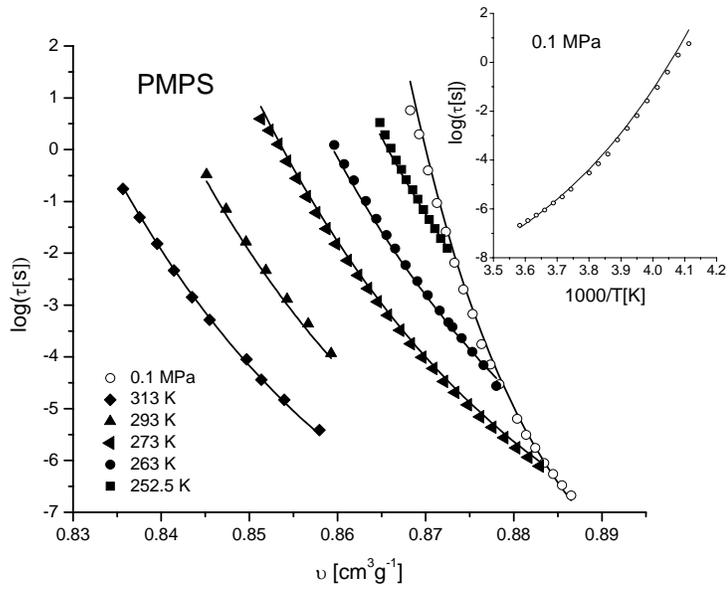

Figure 6

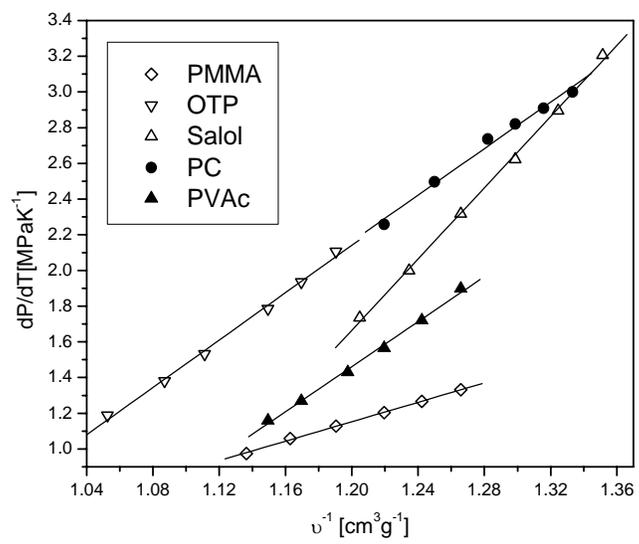

Figure 7